# Stacking control in graphene-based materials: a promising way for fascinating physical properties

Jiliang Zhang,[1,*] Guangcun Shan,[2,3,*]

[1]Department of Energy & Materials Engineering, Dongguk University, Seoul, 04620 Republic of Korea
[2]Department of Materials Science and Engineering, City University of Hong Kong, Kowloon, Hong Kong SAR, China
[3]Institute of Quantum Sensing, School of Instrumentation Science and Opto-electronic Engineering, Beihang University, Beijing, China

**Abstract:**
Graphene, defined as a single atomic plane of graphite, is a semimetal with small overlap between the valence and the conduction bands. The stacking of graphene up to several atomic layers can produce diverse physical properties, depending on the stacking way. The bilayer graphene is also a semimetal, adopting the AB-stacked (or Bernal-stacked) structure or the rare AA-stacked structure. The trilayer or a few layer graphene (FLG) can be semimetal or semiconductor, depending on whether it takes Bernal (ABA) stacking or rhombohedral (ABC) stacking. We will give a perspective on the recent two mild approaches to control the stacking via local transition from ABC stacking into ABA stacking. It is believed that with the rapid development of graphene-based materials, these techniques for stacking control can be used for more complex structure to fulfill fascinating properties and devices.

Graphene, defined as a single atomic plane of graphite, is a semimetal with small overlap between the valence and the conduction bands [1]. The stacking of graphene up to several atomic layers can produce diverse physical properties, depending on the stacking way. The bilayer graphene is also a semimetal, adopting the AB-stacked (or Bernal-stacked) structure or the rare AA-stacked structure [2]. The trilayer or a few layer graphene (FLG) can be semimetal or semiconductor, depending on whether it takes Bernal (ABA) stacking or rhombohedral (ABC) stacking.

Compared with these ideal stacking structure, the stacking structure in most graphene multilayers is actually more complex. Twinning is a common feature, where different stacking configurations coexist, connected by discrete twin boundaries [3]. Twisted multilayers, where one layer is rotated relative to another, have also been observed experimentally. These complex structures usually lead to more fruitful or even exotic physical properties. It has recently been reported in 2018 that a small twist angle of 1.1° induced unconventional superconductivity [4], and this feature can be used for a superconducting transistor made of graphene for quantum devices.

---

[*] Email: jiliangz@dongguk.edu, gshan2@cityu.edu.hk

Owing to the possibility of different stacking configurations, controlling the stacking structure of graphene materials is vital to realize specific properties and/or novel functional devices. In early 2018, Japanese scientists reported their success in selective fabrication of pure ABA and ABC trilayer graphene [5]. Their methods include a heat treat at high temperatures up to 1510 °C under pressure or high vacuum (ca. $10^{-7}$ Torr), and the precise temperature and pressure control is the key to the selective fabrication [5]. Obviously, the application of their methods will be limited by these harsh conditions.

Very recently, Latychevskaia *et al.* [6] has presented two mild approaches to control the stacking via local transition from ABC stacking into ABA stacking. One approach is joule heating by passing high currents through a hBN/ABC FLG/hBN heterostructures, where hBN is the hexagonal boron nitride, and the fraction of the ABA stacking can be controlled by tuning the applied voltage. This approach has a spatial resolution of ca. 1um for the initial transition, which could be utilized as a more effective way to fabricate many graphene-based electronic devices with different stacking structures. The second approach is illuminating by laser pulses to trigger the transition in the laser focused zone.

In joule heating, the high temperature induced stress was thought to be the driving force for the transition. However, the high density direct current can also induce significant stress or structural change due to strong electron-lattice couplings [7,8]. It would be more promising to explore the possibility of the large current to tune the stacking. An electropulsing technique [8], which employs a high density direct current and generates less heat, can be employed for this purpose. With the high density electric pulses, it is possible to bind two separate FLGs and produce twisted layers or numerous interesting stacking configurations.

Laser pulses of 790 nm wavelength were used to initialize the transition in a focused zone of ca. 20 um as the second method. The stacking transition is triggered by the laser illumination the focused zone. Compared with the laser, a short-wavelength UV light (10 nm- 400nm), which generates much higher energy, is more promising to control the stacking transition in a smaller focused zone. This method will allow customized design of the stacking structure on a scale of nanometers by controlling the pathway of the pulsing beam like stereo lithography, which widen the feasibility of making novel graphene-based functional devices.

Although nanospot angle-resolved photoemission spectroscopy is powerful to characterize the local stacking structure [9], many researchers cannot access it due to the limited availability of such synchrotron facilities. Latychevskaia and coworkers have demonstrated that a convergent beam electron holographic technique allows the characterization and reconstruction of the stacking and defects (e.g. misorientation, strain, ripples etc.) at a subnanometer spatial resolution [10,11]. The qualitative information on these defects can also be extracted from obtained patterns or images via using the technique, which enable the further optimization of controlled stacking structures.

The above techniques to control and characterize the stacking structure can also be applied to other 2D or layered materials, which makes it possible to introduce more exciting properties or solve structural mysteries. One example is graphitic $C_3N_4$ (g-$C_3N_4$), the most stable allotrope of carbon nitrides and a fascinating 2D semiconductor. In past decades, many efforts to make g-$C_3N_4$ produce mainly graphitic s-heptazine based $C_3N_4$ (gh-$C_3N_4$), containing a small amount of H [12]. The successful synthesis of graphitic s-trizaine based $C_3N_4$ (gt-$C_3N_4$), the final condensation form of gh-$C_3N_4$, was reported in 2014, and only triple layers are obtainable by exfoliation in the work [13]. Despite these efforts, the 3D stacking structure of these g-$C_3N_4$ is still unsolved. Now it is possible to uncover and modify the stacking, which will open new possibilities for post-silicon electronic devices.

With the rapid development of graphene-based materials, these techniques for stacking control can be used for more complex structure to fulfill fascinating properties and devices. However, challenges still lie ahead, e.g. encapsulation of heterostructure, precise control of heating effects, mechanical performance of devices, etc. To tackle these challenges, in-depth understanding is required not only on these techniques but also on the structure-properties relationships of graphene-based materials, which also leads to better structural design. Compared with the extensive studies in structure and electronic behaviors, less attention was paid to mechanical properties of graphene-based materials and functional devices. From the perspective of practical application, the mechanical properties are also very important and therefore significant efforts are needed to be devoted [14]. Theoretical investigations based on first-principle calculations will provide a simple and effective way to design the desirable structure for verifying the performance on demand [15], and meanwhile utilize the multi-physics simulations further to optimize the technical parameters [16]. It is worthwhile pointing out that it is a matter of time to overcome these challenges. It can be postulated that more amazing 2D graphene-related materials with well-controlled van der Waals 2D heterostructures will be available soon, which endows not only extremely exciting physical phenomena but also superior functional devices for various applications.